\documentclass[conference,comsoc]{IEEEtran}
\usepackage[T1]{fontenc}

\usepackage{amssymb,amsmath,amsfonts,amsthm}
\usepackage{mathrsfs}
\usepackage{cite}
\usepackage{array}
\usepackage{tabularx}
\usepackage[table]{xcolor}
\usepackage{color}
\usepackage{mathtools}
\usepackage{subfigure}
\usepackage{mathtools}
\usepackage{graphicx}
\usepackage{bbm}
\usepackage[utf8]{inputenc}
\usepackage{algorithm}

\newtheorem{lemma}{Lemma}

\newcommand{\maximize}{\mathop{\rm maximize}\limits}

\IEEEoverridecommandlockouts
\ifCLASSINFOpdf
\else
\fi
\usepackage[cmintegrals]{newtxmath}
\begin{document}
	\title{Dynamic Scheduling and Power Control in Uplink Massive MIMO with Random Data Arrivals}
	
	\author{Zheng~Chen, Emil~Bj\"{o}rnson, and~Erik~G.~Larsson\\
		Department of Electrical Engineering (ISY),  Link\"{o}ping University, Link\"{o}ping, Sweden\\
		  Email: \{zheng.chen, emil.bjornson, erik.g.larsson\}@liu.se.
		\thanks{This work was supported in part by ELLIIT, CENIIT, and the Swedish Foundation for Strategic Research.}}
	\maketitle

\begin{abstract}	
In this paper, we study the joint power control and scheduling in uplink massive multiple-input multiple-output (MIMO) systems with random data arrivals. The data is generated at each user according to an individual stochastic process. Using Lyapunov optimization techniques, we develop a dynamic scheduling algorithm (DSA), which decides at each time slot the amount of data to admit to the transmission queues and the transmission rates over the wireless channel. The proposed algorithm
achieves nearly optimal performance on the long-term user throughput under various fairness policies. Simulation results show that the DSA can improve the time-average delay performance compared to the state-of-the-art power control schemes developed for Massive MIMO with infinite backlogs.
\end{abstract}

\section{Introduction}
Massive multiple-input multiple-output (MIMO) is one of the key technologies in 5G \cite{mmimo_mag,Parkvall2017a}. By deploying base stations (BSs) equipped with many antennas, spatial multiplexing can be utilized to serve a large number of users on the same time-frequency resource. The rates and energy efficiency of the network can be largely improved by Massive MIMO compared to the conventional MIMO systems \cite{marzetta2016fundamentals,massivemimobook}. 
Power control is a critical aspect of Massive MIMO systems and it has been extensively studied in many different scenarios \cite{massivemimobook,power_control,Guo2014a}, under the common assumption of an infinite backlog, i.e., there is an infinite amount of data waiting to be transmitted. Since the ergodic rates of both correlated and uncorrelated fading channels can be obtained in a tractable form in the backlogged case, the power control can be optimized with respect to the long-term rate performance, instead of changing with the small-scale fading realizations as was previously the common practice \cite{Bjornson2016b}. 
Since the wireless data traffic usually arrives in a random and bursty manner, 
the set of active users will change dynamically over time. Considering a Massive MIMO system with transmission queues that contain data to be transmitted over the wireless channel, the burstiness of data traffic becomes an important factor for the optimal resource allocation, power control and scheduling policy.

\subsection{Related Work}
The basis of stochastic network optimization was presented in \cite{neely2010stochastic}, with many examples of its applications to communication and queueing systems. Flow control algorithms in a heterogeneous network using Lyapunov optimization techniques were proposed in \cite{fairness_neely}. Furthermore, \cite{georgiadis2006resource} gives an overview of cross-layer control and scheduling algorithms using the drift-plus-penalty framework to achieve stability while optimizing some network performance metric.
MIMO downlink scheduling using the flow control algorithm was studied in \cite{scheduling}, where the backlog at the BS is assumed to be infinite. Due to the difficulty of solving the weighted sum rate maximization problem, an on-off scheduling policy was considered as an approximation of the optimal rate allocation problem.
A dynamic scheduling scheme using Lyapunov techniques was developed in \cite{urllc} for
millimeter wave-enabled Massive MIMO systems, after some simplifying assumptions on the rate expressions. Note that \cite{power_control} proposes an effective algorithm to solve the weighted sum rate maximization problem for Massive MIMO systems with i.i.d. Rayleigh fading channels, which facilitate the application of Lyapunov optimization in Massive MIMO with random data traffic. 

\subsection{Contributions}
In this work, we study the performance of flow control and dynamic rate allocation scheme in uplink Massive MIMO networks with randomly generated data traffic. Different from the scenario with infinite backlog, where the power control is determined and fixed during the entire data transmission period, we develop a dynamic scheduling algorithm (DSA) that decides at each time slot the amount of data that can be admitted to the transmission queues and allocates the appropriate transmission rates to each user. Using Lyapunov optimization theory, our dynamic control policy stabilizes all the transmission queues, while maximizing some concave non-decreasing fairness function on the long-term user throughput. We show that even though the optimization is based on throughput-related utility, the DSA can greatly reduce the time-average delay of the network.

\section{Network Model}
We consider a single-cell uplink massive MIMO network where a base station (BS) with $M$ antennas serves $K$ single-antenna users simultaneously. We assume that the transmission time of the physical layer (PHY) data is divided into fixed-size slots, where each slot contains the transmission time of one or multiple PHY frames.
At each time slot $t\in \{0,1,2,\ldots\}$, uplink data packets from user $k$ are generated according to a stationary and ergodic stochastic process $B_{k}(t)$ and the data generation/arrival rate of this user is $\lambda_k=\mathbb{E}[B_{k}(t)]$ bit/slot. The packet-generating processes for the $K$ users are independent of each other.
The generated data is stored in the transport layer reservoir, which is assumed to have infinite size. We assume that each user maintains a transmission queue at the data link layer, which contains the data ready to be transmitted over the wireless channel to the BS \cite{queue_adaptive}. Denote by $L_k(t)$ and $Q_k(t)$  the current backlog length (in bits) of user $k$ at slot $t$. To avoid congestion in the transmission queues, only a fraction of the data in the transport layer reservoir is allowed to enter the transmission queue.\footnote{The separation between transport layer reservoir and PHY transmission queues also helps us to use Lyapunov optimization techniques to design a dynamic control policy, since this method requires stable queues.} The amount of admitted data at each slot $t$ is denoted by $A_k(t)$ with $r_k=\lim\limits_{t\rightarrow\infty}\frac{1}{t}\sum\limits_{\tau=0}^{t-1}\mathbb{E}[A_k(\tau)]$ bit/slot being the time-average admitted date rate of user $k$. 
Due to the random data arrivals, the admitted data to the transmission queues at every slot $t$ must not exceed the total amount of data in the reservoir: $A_k(t)\leq L_k(t)$.

Denote by $R_k(t)$ the PHY transmission rate of uplink user $k$ in slot $t$, measured by the number of bits that can be delivered over the wireless channel to the BS. The transmission queue $Q_k(t)$ is updated by the following equation:
\begin{equation}
Q_k(t+1)=\max[Q_k(t)-R_k(t),0]+A_k(t), \quad \forall k.
\label{eq:evolution_Q}
\end{equation}
Here, the transmission rates $R_k(t)$ are limited by the network topology, the channel statistics, and the power constraint.

\begin{figure}[t!]
	\centering
	\includegraphics[width=0.95\columnwidth]{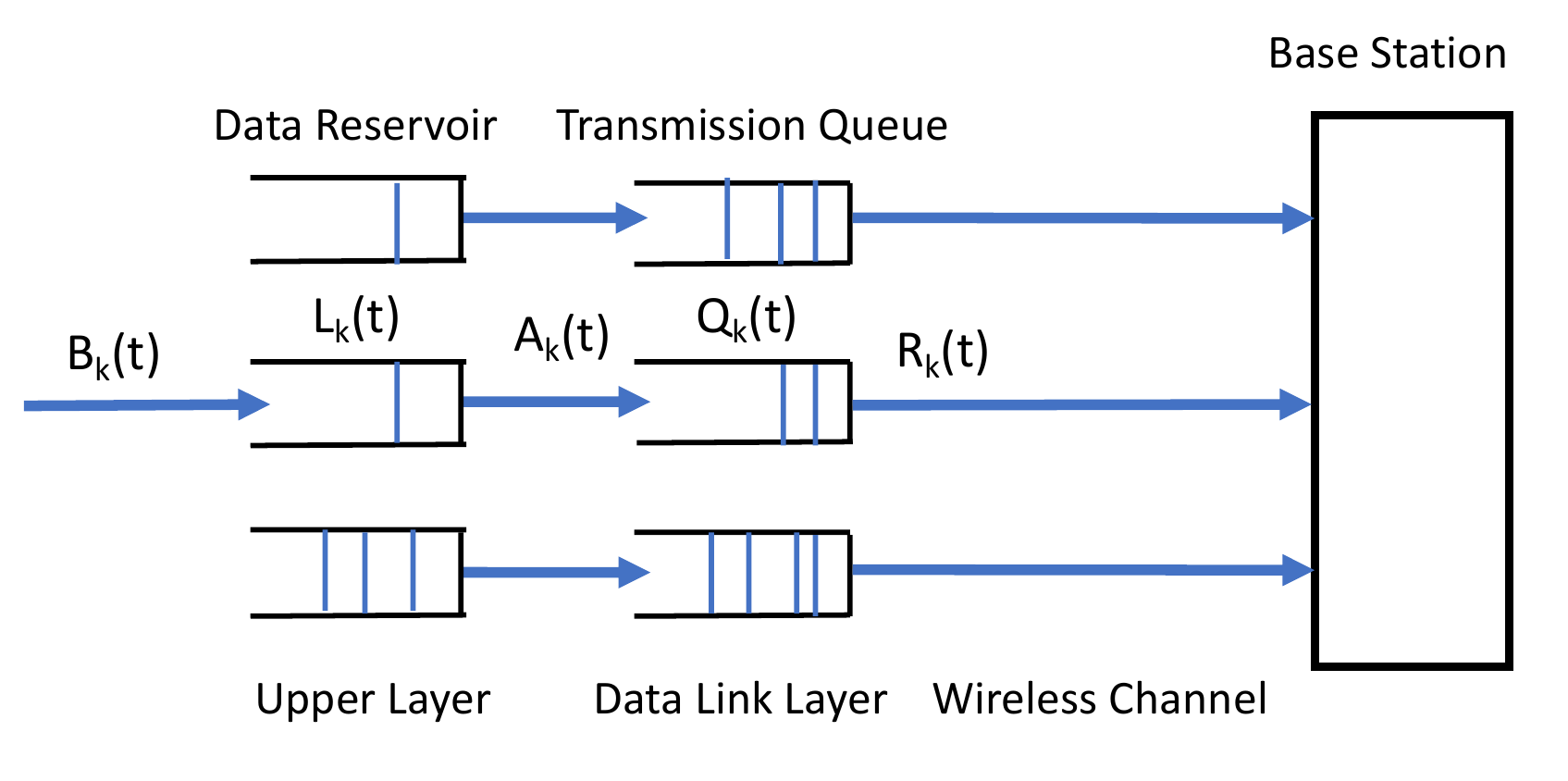}
		\vspace{-0.3cm}
	\caption{The structure of uplink Massive MIMO system, which consists of the data backlog reservoir and the transmission queues. }
	
	\label{fig:system}
	\vspace{-0.2cm}
\end{figure}

\begin{figure}[ht!]
	\centering 
	\includegraphics[width=0.9\columnwidth]{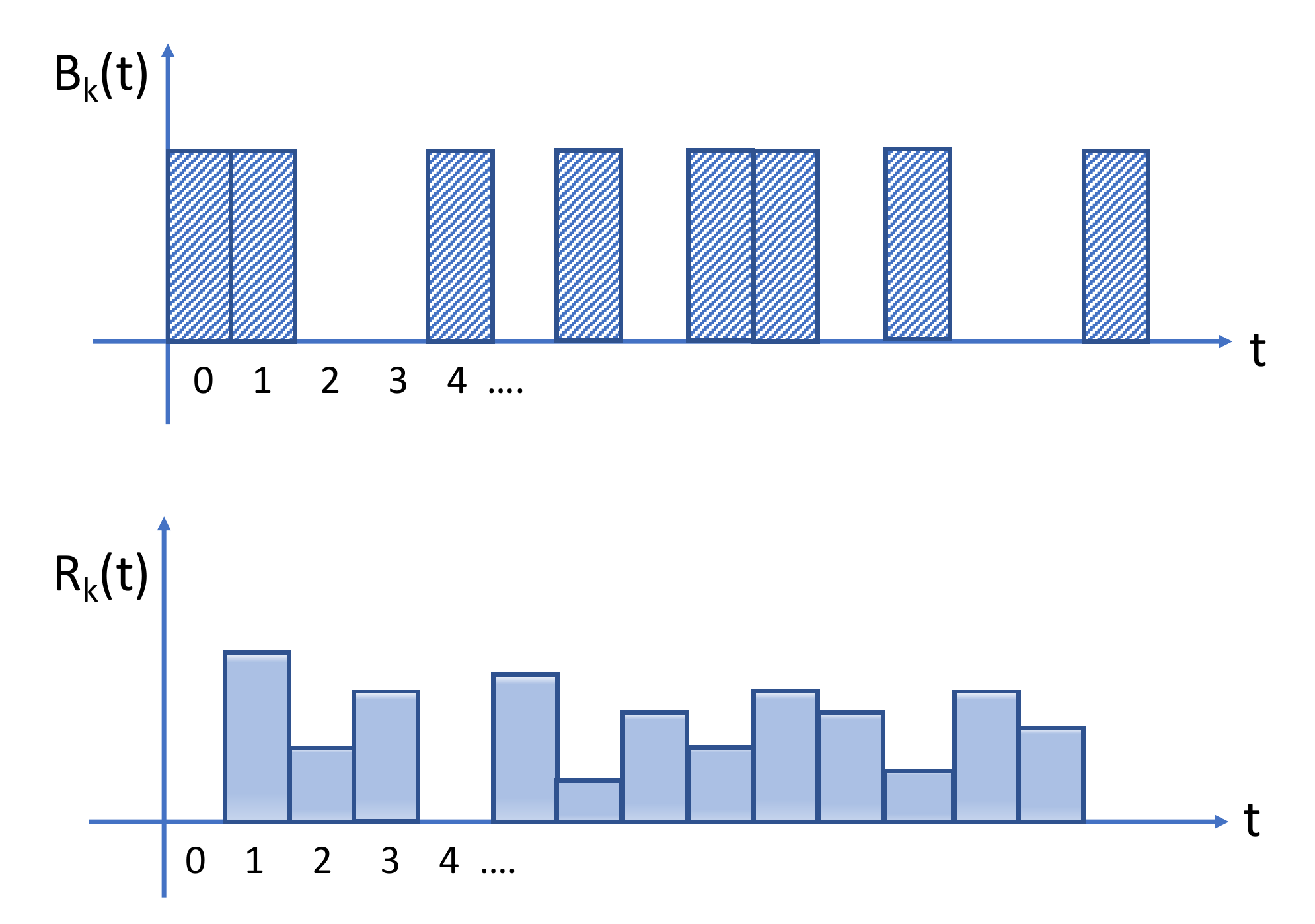} 
		\vspace{-0.2cm}
	\caption{An illustration example of data arrival process and PHY transmission. At each slot, a certain amount of data is generated at each user with some probability. As a result of the dynamic power control, the PHY transmission rates might vary in different slots.}
	\label{fig:arrival_transmission}
			\vspace{-0.3cm}
\end{figure}

For notational convenience, we define the queue vectors $\mathbf{Q}(t)=[Q_1(t),\ldots, Q_K(t)]$, $\mathbf{R}(t)=[R_1(t),\ldots, R_K(t)]$, $\mathbf{A}(t)=[A_1(t),\ldots, A_K(t)]$ and $\mathbf{L}(t)=[L_1(t),\ldots, L_K(t)]$.
The system model is shown in Fig.~\ref{fig:system}. The evolution of the data arrival and PHY transmission processes is illustrated in Fig.~\ref{fig:arrival_transmission}.

\subsection{Problem Formulation}
Due to the random data arrivals, the $K$ uplink users in the network will not always have data to transmit.
For arbitrary data arrival rates, our objective is to develop a dynamic control policy that:
\begin{enumerate}
	\item maintain the transmission queues stable;
	\item achieve a long-term throughput vector that maximizes some utility function $f(\cdot)$.
\end{enumerate}
In this dynamic control problem, at every slot $t$, we need to decide the amount of data to be admitted to the transmission queues, and perform power control which determines the transmission rates allocated to each user. Thus, the control decisions are $\alpha(t)=[\mathbf{A}(t); \mathbf{P}(t)]$, where $\mathbf{P}(t)=[P_1(t), \ldots, P_K(t)]$ is the  power control vector. Let $\mathcal{A}(t)$ denote the set of all possible control actions in slot $t$, given the random data arrivals and power constrains in this slot.

When the transmission queues are stable, the long-term throughput vector is equal to the time average admitted data rate vector $\boldsymbol{r}=[r_1, \ldots, r_K]$. We define $\overline{X}=\lim\limits_{t\rightarrow\infty}\frac{1}{t}\sum\limits_{\tau=0}^{t-1}\mathbb{E}\{X(t)\}$ as the time-average of a random process $X(t)$. The utility maximization is thus defined by the following problem:
\begin{subequations}
	\begin{align}
	\maximize~~&f(\boldsymbol{r}) \label{eq:optimization_utility}\\
	\textrm{subject~to}~~&0\leq r_{k}\leq \lambda_k, \quad \forall k \label{condi2}\\
	&\overline{Q}_k< \infty, \quad \forall k \label{condi3}\\
	&\alpha(t)\in\mathcal{A}(t), \quad \forall t.
	\end{align}
	\label{eq:optimization-prob}
\end{subequations}
Here, the network utility function $f(\cdot)$ needs to be an element-wise non-decreasing concave function. It can reflect one out of the many fairness criteria that will be presented in Section \ref{sec:utility}.
The condition in \eqref{condi2} ensures that the time-average throughput of user $k$ is not larger than the generated uplink data rate of this user. \eqref{condi3} is the strong stability condition of the transmission queues.

\subsection{Ergodic Rates in Massive MIMO}
In this work, we consider block fading channels with i.i.d.~Rayleigh fading between the $M$ BS antennas and the $K$ single-antenna users. This assumption allows us to derive a simple yet rigorous lower bound on the achievable ergodic rates of the users, which only depend on the large-scale fading parameters and the power control scheme. Though in reality the channels are unlikely to be i.i.d.~Rayleigh fading \cite{massivemimobook}, it has been shown in \cite{measured} that the achievable rates obtained by real measured channels are close to the rates obtained by assuming i.i.d.~Rayleigh channels. This is probably a consequence of the channel hardening and favorable propagation properties of Massive MIMO, which make the rates less dependent on the actual channel distributions, and mainly a functions of the average pathlosses.

The maximum achievable ergodic uplink rate of user $k$, measured in bit/slot, is lower-bounded by \cite{power_control}\footnote{To achieve the ergodic rate, we need to transmit codewords that span many channel realizations. In practice, this means transmitting at least 1\,kB of data \cite{Bjornson2016b}, which is easily done over a short time slot by using many sub-carriers.}
\begin{equation}
R_k=(\tau_c-\tau_p)\log_2(1+\text{SINR}_k),
\label{eq:rate_bit}
\end{equation}
where $\tau_c$ is the length of the coherence block and $\tau_p\in[K, \tau_c]$ denotes the length of the pilot signal. With maximum-ratio combining (MRC), we have
\begin{equation}
\text{SINR}_k=\frac{M p_{\text{d},k}\gamma_{k}}{1+\sum_{j=1}^{K}\beta_{j}p_{\text{d},j}}, 
\label{eq:mrc}
\end{equation}
where $\beta_k$ is the large-scale fading coefficient of user $k$, including the pathloss and shadowing; $\gamma_k=\frac{\tau_p p_{\text{p},k}\beta_{k}^2}{1+\tau_p p_{\text{p},k}\beta_{k}}$ is the mean square of the channel estimates; $p_{\text{p},k}$ and $p_{\text{d},k}$ denote the pilot and payload power levels, respectively. Since the transmission rate is not necessarily an integer, we assume that the data can be admitted and transmitted as fractional frames.

Note that the scheduling policy considered in this work also works for the ZF detection, after replacing \eqref{eq:mrc} with the respective SINR expression, but we limit the study to MRC due to the space limitation. More details on the ergodic rates and their derivation can be found in \cite{marzetta2016fundamentals,massivemimobook}.

\section{Lyapunov Optimization and Dynamic Joint Scheduling and Power Control Algorithm}
The optimization problem in \eqref{eq:optimization-prob} involves maximizing functions of a time-average quantity. In order to develop a dynamic joint scheduling and power control algorithm that achieves performance arbitrarily close to the optimal solution, we use the Lyapunov optimization theory with the help of virtual queues.\footnote{Note that if the utility function $f(\cdot)$ is linear, maximizing $\mathbb{E}[f(\boldsymbol{A}(t))]$ is equivalent to maximizing $f(\boldsymbol{r})$. We can use a CLC1-type algorithm as proposed in \cite{fairness_neely} to achieve stability with performance optimization. For general utility function which is not necessarily linear, maximizing function of time-average utility requires the usage of virtual queues, as explained in \cite{neely2010stochastic}.}
We introduce auxiliary variables $\boldsymbol{\nu}(t)=[\nu_{1}(t),\ldots,\nu_{K}(t)]$  for each admitted data stream $A_{k}(t)$ and the corresponding virtual queues $\mathbf{Y}(t)=[Y_1(t),\ldots, Y_K(t)]$ that evolves as follows:
\begin{equation}
Y_k(t+1)=\max[Y_k(t)-A_k(t),0]+\nu_{k}(t).
\label{eq:evolution_Y}
\end{equation}
Here, $\nu_{k}(t)$ and $Y_k(t)$ are also measured in bits.

\begin{lemma}
The original problem in \eqref{eq:optimization-prob} can be transformed into the following problem that involves maximizing the time-average of the network utility:
\begin{subequations}
	\begin{align}
	\maximize~~&\overline{f(\boldsymbol{\nu})}\\
	\textnormal{subject~to}~~& \overline{\nu}_k\leq r_k, \quad \forall k \label{cons1}\\
	&\overline{Q}_k<\infty, \quad \forall k\\
	& 0\leq \nu_k(t)\leq A_{\max}, \quad \forall k, t \label{cons2}\\
	&\alpha(t)\in\mathcal{A}(t), \quad \forall t.
	\end{align}
	\label{eq:optimization-prob2}
\end{subequations} $\overline{f(\boldsymbol{\nu})}=\lim\limits_{t\rightarrow\infty}\frac{1}{t}\sum\limits_{\tau=0}^{t-1}\mathbb{E}[f(\boldsymbol{\nu}(t))]$ and $\overline{\nu}_k=\lim\limits_{t\rightarrow\infty}\frac{1}{t}\sum\limits_{\tau=0}^{t-1}\mathbb{E}[\nu_k(\tau)]$ 
are the time-average of the utility function and the arrival rate of the virtual queue, respectively. $A_{\max}$ serves as an upper bound for the auxiliary variables and it is chosen to be suitably large such that $B_k(t)\leq A_{\max}$ always holds.
\end{lemma}
\begin{IEEEproof}
The proof to show that these two problems are equivalent can be found in \cite{neely2010stochastic} and \cite{georgiadis2006resource}.
\end{IEEEproof}

Let $\boldsymbol{\Theta}(t)=[\mathbf{Y}(t);\mathbf{Q}(t)]$ denote the matrix of virtual queues and transmission queues. We consider the following Lyapunov function
\begin{equation}
\mathcal{L}(\boldsymbol{\Theta}(t))=\frac{1}{2}\sum\limits_{k=1}^{K}Q_k^2(t)+\frac{\eta}{2}\sum\limits_{k=1}^{K}Y_k^2(t),
\vspace{-0.1cm}
\label{eq:lyapounov-function}
\end{equation}
where $0<\eta\leq 1$ is a bias factor that determines the relative weight on the virtual queues. The Lyapunov function is a scalar measure of the congestion level in the system.
The one-step conditional Lyapunov drift is
\begin{equation}
\Delta\big(\boldsymbol{\Theta}(t)\big)=\mathbb{E}[\mathcal{L}(\boldsymbol{\Theta}(t+1))-\mathcal{L}(\boldsymbol{\Theta}(t))|\boldsymbol{\Theta}(t)].
\vspace{-0.1cm}
\label{eq:one-step-drift}
\end{equation}  
To maximize the time-average of the network utility while stabilizing the combined queues $\boldsymbol{\Theta}(t)$, we use the min-drift-plus-penalty technique introduced in \cite{neely2010stochastic}. At every time slot, we aim at designing a dynamic control policy that minimizes the drift-plus-penalty expression:
\begin{equation}
\Delta(\boldsymbol{\Theta}(t))-V \mathbb{E}[f(\boldsymbol{\nu}(t))|\boldsymbol{\Theta}(t)],
\end{equation}
where $V>0$ is a control parameter that leverages between network utility and the congestion in the queues.

\begin{lemma}
	\label{lemma1}
The drift-plus-penalty is upper bounded as 
\begin{align}
&\Delta(\boldsymbol{\Theta}(t))-V \mathbb{E}[f(\boldsymbol{\nu}(t))|\boldsymbol{\Theta}(t)]\nonumber\\
&\leq  
C-\mathbb{E}\left[\sum_{k=1}^{K}A_k(t)\big(\eta Y_k(t)-Q_k(t)\big)|\boldsymbol{\Theta}(t)\right]  \nonumber\\
&-\mathbb{E}\left[Vf(\boldsymbol{\nu}(t))\!-\!\eta \sum\limits_{k=1}^{K}Y_k(t)\nu_k(t)|\boldsymbol{\Theta}(t)\right]\nonumber\\
&-\mathbb{E}\left[\sum_{k=1}^{K}Q_k(t) R_k(t)|\boldsymbol{\Theta}(t)\right],
\label{eq:drift-penalty}
\end{align}
where $C=\frac{1}{2}\sum\limits_{k=1}^{K}R_{k,\max}^2+\frac{2\eta+1}{2}\sum\limits_{k=1}^{K}\lambda_k^2.$
\end{lemma}
\begin{IEEEproof}
	See Appendix \ref{appen1}.
\end{IEEEproof}

Minimizing the right-hand side of the drift-plus-penalty expression in \eqref{eq:drift-penalty} at each slot $t$ leads to the need to solve three subproblems, which are solved in the following subsections.

\subsection{First Subproblem: Data Admission Control}
To minimize the first non-constant term of the right-hand side of \eqref{eq:drift-penalty}, we need to choose $A_k(t)$ that maximizes $\sum_{k=1}^{K}A_k(t)\big(\eta Y_k(t)-Q_k(t)\big)$ under the backlog constraint $A_k(t)\leq L_k(t)$ and the admission burst limit $A_k(t)\leq A_{\max}$. The solution to the first subproblem is
	\begin{equation}
	A_k(t)= 
	\left\lbrace 
	\begin{array}{ccc}
	\!\!\min\{L_k(t), A_{\max}\}
	& \text{if}~Q_k(t)\leq\eta Y_k(t), \\
	0
	& \text{otherwise.}
	\end{array} \right.
	\end{equation}

\subsection{Second Subproblem: Auxiliary Variables}	
\vspace{-0.1cm}
\label{sec:utility}
To minimize the second non-constant term of the right-hand side of \eqref{eq:drift-penalty}, we need to choose the auxiliary variables $0\leq \nu_k(t)\leq A_{\max}$ that solve\vspace{-0.3cm}
\begin{equation}
\maximize\limits_{\substack{0\leq\nu_k(t)\leq A_{\max}\\\boldsymbol{\nu}(t)=[\nu_1(t),\ldots,\nu_{K}(t)]}}~Vf(\boldsymbol{\nu}(t))-\eta \sum\limits_{k=1}^{K}Y_k(t)\nu_k(t). 
\vspace{-0.1cm}
\label{eq:utility-maximization}
\end{equation}The solution to this problem depends on the specific utility function that we choose. Two examples are given below. 

\subsubsection{Max-Min Fairness (MMF)}
In this case, every user should achieve the same performance, thus we have the utility function 
\vspace{-0.3cm}
\begin{equation}
f(\boldsymbol{\nu}(t))=\min\{\nu_1(t), \nu_2(t),\ldots, \nu_K(t)\}.
\vspace{-0.1cm}
\end{equation}
The solution to \eqref{eq:utility-maximization} is the case when all $\nu_k(t)$ are the same and when $\nu_k(t)[V-\eta \sum\limits_{j=1}^{K}Y_j(t)]$ is maximized. Combined with $0\leq \nu_k(t)\leq A_{\max}$, we have the solution to \eqref{eq:utility-maximization} as\vspace{-0.1cm}
\begin{align}
\nu_k(t)= \left\{
\begin{array}{rcl}
&A_{\max} & \text{if}~~V>\eta\sum_{j=1}^{K}Y_j(t),\\
&0 & \text{otherwise.}\\
\end{array} \right.
\end{align}

\vspace{-0.1cm}
\subsubsection{Maximum Sum Rate (MSR)}
In this case, the system should maximize the data throughput without taking fairness between users into consideration. We then have the utility function\vspace{-0.15cm}
\begin{equation}
f(\boldsymbol{\nu}(t))=\sum_{k=1}^{K} (\nu_k(t)).
\vspace{-0.1cm}
\end{equation}
The solution to \eqref{eq:utility-maximization} is\vspace{-0.2cm}
\begin{align}
\nu_k(t)= \left\{
\begin{array}{rcl}
&A_{\max} & \text{if}~~V>\eta Y_k(t),\\
&0 & \text{otherwise.}\\
\end{array} \right.
\end{align}

Depending on the specific form of the utility function $f(\cdot)$, closed-form solutions to \eqref{eq:utility-maximization} might not be available, but standard convex solvers can be used to find the optimal solutions.

\vspace{-0.1cm}
\subsection{Third Subproblem: PHY Rate Allocation} 
To minimize the third non-constant term of the right-hand side of \eqref{eq:drift-penalty}, we choose $R_k(t)$ that maximizes $\sum_{k=1}^{K}Q_k(t) R_k(t)$ given the uplink power constraints $0\leq P_k(t) \leq P_{\max}$, $\forall k, t$, which we identify as a weighted sum rate problem.
This is often the most challenging subproblem since the maximization of the weighted sum rate is an NP hard problem in many cases \cite{Luo2008a}. However, recently \cite{power_control} developed an efficient algorithm that exploits the special structure of the rates in Massive MIMO to solve this weighted sum rate optimization problem. In Section~\ref{sec:simulation}, our simulation results are obtained with the help of the algorithm developed in \cite{power_control}.

In summary, we have developed the novel dynamic scheduling algorithm (DSA) that is given in Algorithm~\ref{algorithm:DSA}, where one iteration is taken per time slot.
 
\begin{algorithm}
	\caption{Dynamic Scheduling Algorithm (DSA)} \label{algorithm:DSA}
\begin{enumerate}
	\item Initialization: $L_k(0)=0$, $Q_k(0)=0$ and $Y_k(0)=0$ for all $k=1,\ldots,K$. Set $t=1$.
	\item At current slot $t$, obtain the input vector $\boldsymbol{\nu}(t)$ at the virtual queues by solving 
	\vspace{-0.3cm}
	\begin{equation}
	\maximize\limits_{\substack{0\leq\nu_k(t)\leq A_{\max}\\\boldsymbol{\nu}(t)=[\nu_1(t),\ldots,\nu_{K}(t)]}}\left[V\cdot f\big(\boldsymbol{\nu}(t)\big)-\eta\sum\limits_{k=1}^{K}Y_k(t)\nu_{k}(t)\right].
	\label{eq:solution} \nonumber
	\end{equation} 
	Here, $A_{\max}$ and $V$ are suitably large constant parameters. 
	\item Obtain the admitted data $A_k(t)$ at the transmission queues by\vspace{-0.15cm}
		\begin{equation}
	A_k(t)= 
	\left\lbrace 
	\begin{array}{ccc}
	\!\!\min\{L_k(t), A_{\max}\}
	& \text{if}~Q_k(t)\leq\eta Y_k(t), \\
	0
	& \text{otherwise}.
	\end{array} \right.\nonumber
	\end{equation}
	\item Determine transmission rates $R_k(t)$ by solving the weighted sum rate maximization problem (as in \cite{power_control}):
		\vspace{-0.3cm}
	\begin{equation}
	\maximize\limits_{\mathbf{R}(t)}~\sum_{k=1}^{K} Q_k(t)\cdot R_k(t). \label{eq:scheduler} \nonumber
	\vspace{-0.15cm}
	\end{equation}
	\item Update the virtual queues, transmission queues and the transport layer reservoirs as follows:
	\vspace{-0.2cm}
			\begin{align} \notag
	 Y_k(t+1)&=\max[Y_k(t)-A_k(t),0]+\nu_{k}(t),\\ \notag	\vspace{-0.15cm}
Q_k(t+1)&=\max[Q_k(t)-R_k(t),0]+A_k(t),\\ \notag  \vspace{-0.15cm}
 L_k(t+1)&=\max[L_k(t)-A_k(t),0]+B_k(t).\vspace{-0.15cm}
 			\end{align}
	\item Continue steps 2--5 for the next slot $t+1$.
\end{enumerate}
\end{algorithm}

\vspace{-0.1cm}
\section{Performance Evaluation}
\label{sec:simulation}
In this section, we evaluate the performance of the proposed DSA with different utility functions, including max-min fairness (MMF) and maximum sum rate (MSR). The parameters used in the simulations are summarized below.
\begin{itemize}
	\item The number of antennas is $M=100$, the number of users is $K=10$,  the number of symbols per coherence interval is $\tau_c=100$, which could be achieved by having a coherence bandwidth of $B=100\,$kHz with the coherence time $T_c=1\,$ms. The pilot signal length $\tau_p$ is equal to the number of effectively scheduled users in a time slot.
	\item The payload power budget of user $k$ is $P_{\max}=10^{0.5}\times R^{3.76}$. The pilot power of all users are $P_{\text{d},k}=P_{\max}$. Table \ref{tab:beta} gives an example of the SNR (dB) values (in ascending order) of the $K=10$ users that we use in the simulations.
	\item The data-generating process of user $k$ follows a memoryless Bernoulli process with packet arriving probability $p_k$ per slot, i.e., $B_k(t)\sim \text{Bernoulli}(p_k)\times B_{\max}$, with $B_{\max}=5\times \tau_c$ bits.
	The duration of one slot is the same as the length of a coherence interval.\footnote{For simplicity, we assumed that there is only one coherence interval per slot, but all the results can be readily applied to a system where one time slot contains multiple coherence intervals that are distributed over the frequency domain. The only change that is needed is to multiply $A_{\max}$, $B_{\max},$ and $V$ with the number of coherence intervals per slot.}
	\item  $A_{\max}=20\times \tau_c$, $V=1000\times \tau_c$, $\eta=0.5$. The results are obtained after at least $10000$ slots.
\end{itemize}

\begin{table}[t]
	\vspace{-0.4cm}
	\centering
	\caption{SNR $(\textnormal{dB})$ of $K=10$ users}
	\renewcommand{\arraystretch}{1.2}
	\begin{tabular}{|c|c|c|c|c|}
		\hline   
	  $\text{SNR}_1$&    $\text{SNR}_2$&  $\text{SNR}_3$&  $\text{SNR}_4$& $\text{SNR}_5$ \\
	  \hline
	  $ -0.62$ & $3.27$  &$ 5.4$  &$6.5$  &$ 9.5$   \\
	  \hline
	  $\text{SNR}_6$&    $\text{SNR}_7$&  $\text{SNR}_8$&  $\text{SNR}_9$& $\text{SNR}_{10}$ \\
	  \hline
	   $10$    &$12.8$  &$15.7$ &$17.56$   & $22.36$ \\
		\lasthline
	\end{tabular}
	\label{tab:beta}
	\vspace{-0.5cm}
\end{table}

For comparison with the state-of-the-art power control schemes, we present results obtained by the MMF and MSR optimization algorithms proposed in \cite{power_control} for users with infinite backlog. Since these algorithm are clearly suboptimal whenever a user has an empty queue, we consider two heuristic benchmark schemes to optimize the MMF and MSR.
The first heuristic scheme is referred to as ``modified MMF" and is defined as follows:
\begin{enumerate}
	\item At time slot $t$, if $L_k(t)=0$, then user $k$ is removed from the list of users waiting to be served. 
	\item Apply the MMF power control algorithm from \cite{power_control} on the $K'$ users with non-empty queues such that max-min fairness is achieved among them. 
\end{enumerate}
The second heuristic scheme is referred to as ``modified MSR'', which uses the weighed sum rate maximization algorithm developed in \cite{power_control} after removing users that have empty queues.

\begin{figure}[th!]
	\vspace{-0.3cm}
	\centering
	\includegraphics[width=\columnwidth]{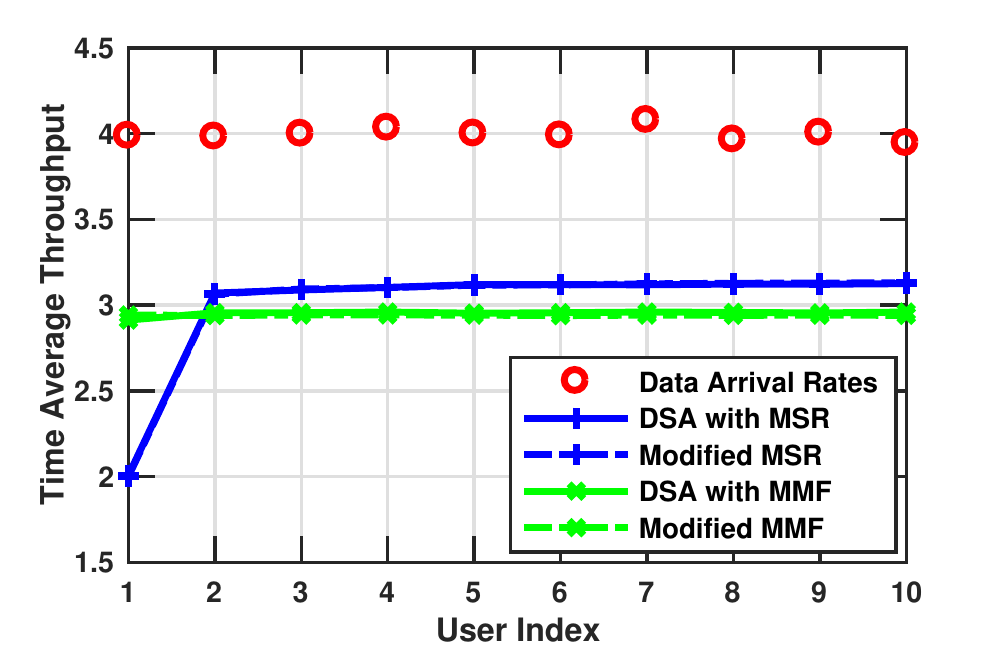}
	\caption{Time-average throughput of $K=10$ users.}
	\label{fig:rate-one-location-same-arrival}
	\vspace{-0.2cm}
\end{figure}

\vspace{-0.15cm}
\subsection{Throughput Comparison}
In Fig.~\ref{fig:rate-one-location-same-arrival}, we compare the time-average throughput (bit/channel use) obtained by the DSA with the two heuristic algorithms. The packet arrival probabilities are $p_k=0.4$ for all $k=1,\ldots,K$. In this case, the data arrival rates exceed the achievable ergodic rates of all users, which means that the backlog goes towards infinity and the users will (almost) always have data to transmit.
Hence, the time-average throughput is limited by the ergodic rates of the users. From Fig.~\ref{fig:rate-one-location-same-arrival}, we see that the two heuristic algorithms provide almost the same throughput as the proposed DSA.
This shows that in terms of throughput optimization, the cross-layer dynamic scheduling using Lyapunov optimization is not needed in this case. On the other hand, the system is not practically useful in this case since the delays go to infinity.  We therefore consider the practical case of having data arrival rates that are inside the ergodic rate region in the remainder of this section.

\begin{figure}[ht!]
	\centering
	\includegraphics[width=\columnwidth]{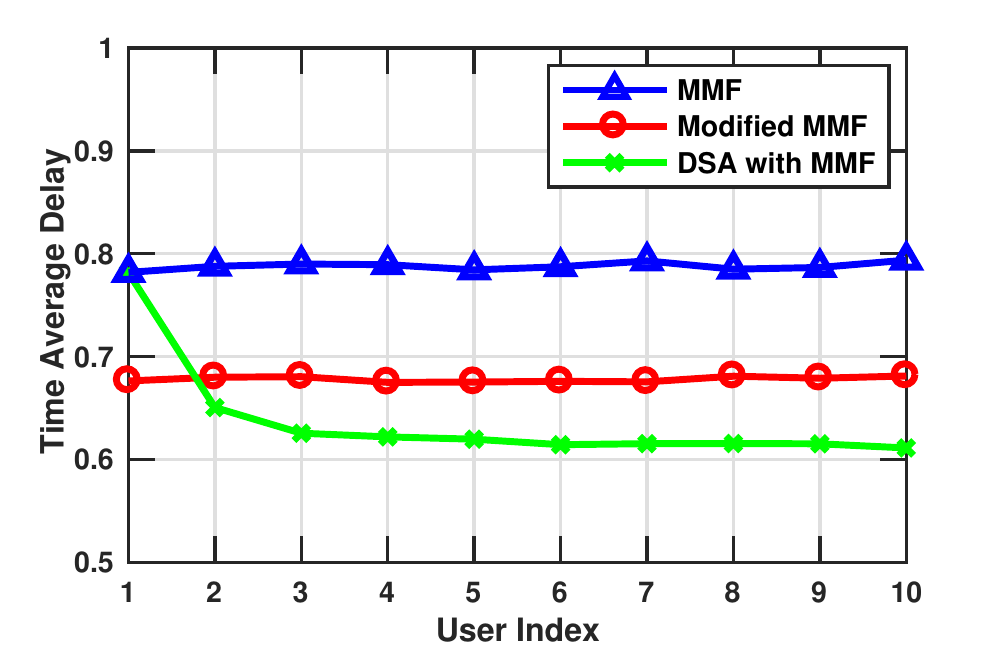}
	\caption{Time-average delay ($\times 10^{-5}\,$s/bit) of $K=10$ users. All the users have the same data arrival rates, i.e., $B_{\max}=5\times\tau_c$, $p_k=0.4$ for all $k=1,\ldots, 10$.}
	\label{fig:delay_comp_mmf}
	\vspace{-0.3cm}
\end{figure}

\begin{figure}[ht!]
	\centering
	\includegraphics[width=\columnwidth]{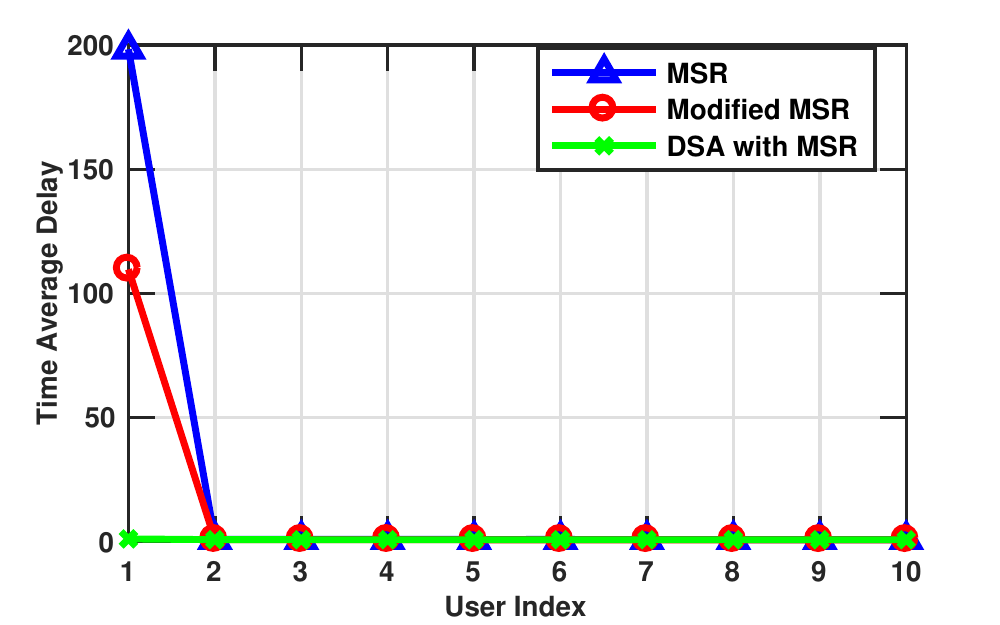}
	\caption{Average delay ($\times 10^{-5}\,$s/bit) of $K=10$ users after $t=10^{4}$ time slots. All the users have the same data arrival rates, i.e., $B_{\max}=5\times\tau_c$, $p_k=0.5$ for all $k=1,\ldots, 10$.}
	\label{fig:delay_comp_msr}
	\vspace{-0.3cm}
\end{figure}

\begin{figure}[ht!]
	\centering
	\includegraphics[width=\columnwidth]{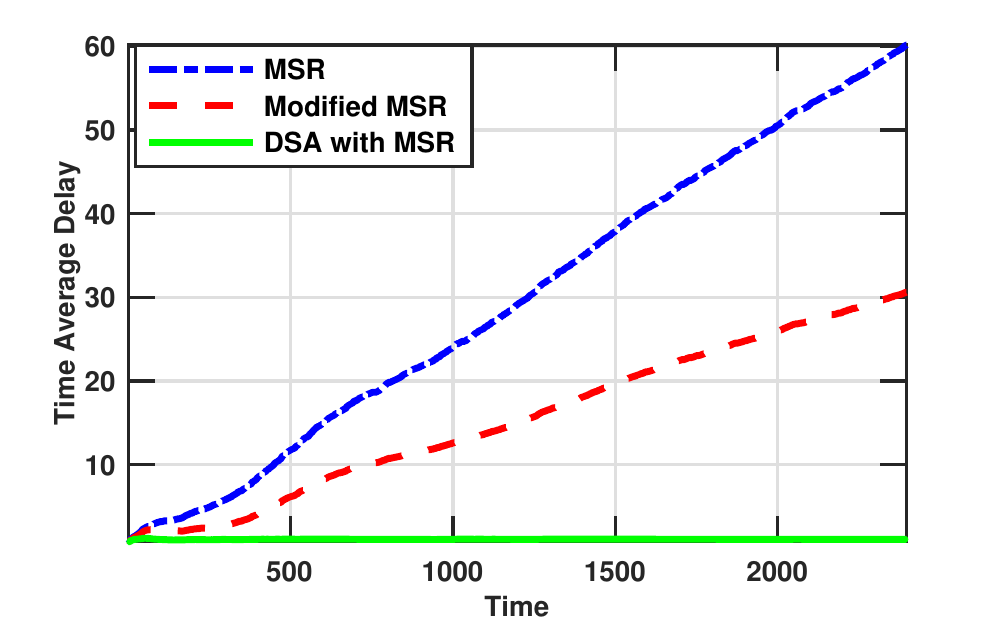}
	\caption{Average delay ($\times 10^{-5}\,$s/bit) of user $1$ vs. time. Same parameters as in Fig.~\ref{fig:delay_comp_msr}.}
	\label{fig:delay_increase}
	\vspace{-0.4cm}
\end{figure}

\vspace{-0.1cm}
\subsection{Delay Comparison}
In Fig.~\ref{fig:delay_comp_mmf}, we consider the time-average delay obtained with the DSA, the original MMF algorithm in \cite{power_control} and the modified MMF. The data arrival probabilities are the same for all users.  We see that the average delays of the $10$ users are very similar when using the MMF and the modified MMF algorithms, because the allocated transmission rates are almost the same. The DSA has a clear advantage in reducing the time-average delay for most users compared to the two alternative algorithms, at the price of giving slightly larger delay for the worst-channel user.

In Fig.~\ref{fig:delay_comp_msr}, we present the delay performance when the objective is to maximize the sum rate. The parameters we choose belong to a particular case where user $1$ has higher data arrival rate than its optimal transmission rate derived with MSR in the infinite backlog case. If we do not take into account the bursty traffic, with conventional MSR scheme, user $1$ will have an unstable queue, which leads to infinite delay with time increasing. However, since the arrival rate vector still falls inside the stable throughput region of our system, the queues can be stabilized when using our DSA. 
From this figure, we can see that the DSA gives small and bounded delay for all users, while for the other two methods, the delay is large and will keep increasing towards infinity as time increases.\footnote{Note that when the queue is unstable, the time-average delay is infinite. The delay presented in Fig.~\ref{fig:delay_comp_msr} is obtained after 10000 time slots, and it will increase to infinity with time.} To demonstrate this, in Fig.~\ref{fig:delay_increase} we show the evolution of the time-average delay with the number of time slots, showing that the delay obtained with the MSR and modified MSR grows linearly with time.

\section{Conclusions}
\vspace{-0.05cm}
In practice, the data arrivals in Massive MIMO systems will be random and this fact makes the resource allocation problem rather different from the infinite backlog scenario that has dominated the literature. 
In this work, we studied the cross-layer flow control and rate allocation in uplink Massive MIMO systems. With the help of Lyapunov optimization theory,  we constructed a dynamic scheduling algorithm that achieves long-term user throughputs by maximizing a predefined utility function. Compared to the conventional deterministic power control schemes, our new algorithm can substantially reduce the average delay experienced by the users and also balance the queues in scenarios where the conventional schemes fail to do that.

\vspace{-0.1cm}
\appendix
\appendices
\subsection{Proof of Lemma \ref{lemma1}}
\label{appen1}
From \eqref{eq:lyapounov-function} and \eqref{eq:one-step-drift}, the Lyapunov drift is computed as
\begin{align}
&\Delta\big(\boldsymbol{\Theta}(t)\big) =\mathbb{E}[\mathcal{L}(\boldsymbol{\Theta}(t+1)-\mathcal{L}(\boldsymbol{\Theta}(t))|\boldsymbol{\Theta}(t)] \nonumber\\
=&\mathbb{E}\left[\frac{1}{2}\sum\limits_{k=1}^{K}\big(Q_k^2(t+1)-Q_k^2(t)\big)+\frac{\eta}{2}\sum\limits_{k=1}^{K}\big(Y_k^2(t+1)-Y_k^2(t)\big)\right].  \nonumber \label{eq:drift2} 
\end{align}
From the queue evolution equations in \eqref{eq:evolution_Q} and \eqref{eq:evolution_Y}, we have
\begin{align}
Q_k(t+1)^2 
\!\leq&\! Q_k(t)^2\!+\!R_k(t)^2\!+\!A_k(t)^2\!-\!2Q_k(t)\big[R_k(t)-A_k(t)], \nonumber\\
Y_k(t+1)^2\leq&  Y_k(t)^2+A_k(t)^2+\nu_k(t)^2-2Y_k(t)\big[A_k(t)-\nu_k(t)].\nonumber
\end{align}
Then we have
\vspace{-0.1cm}
\begin{equation}
\begin{split}
\Delta\big(\boldsymbol{\Theta}(t)\big) \leq&~\frac{1}{2}\sum_{k=1}^{K}\mathbb{E}\left[R_k(t)^2+A_k(t)^2|\boldsymbol{\Theta}(t)\right]\\
\vspace{-0.05cm}&-\!\!\sum_{k=1}^{K}\!\mathbb{E}\Big[Q_k(t)\big(R_k(t)-A_k(t))|\boldsymbol{\Theta}(t)\Big]
\\\vspace{-0.05cm}&\!+\!\frac{\eta}{2}\!\sum_{k=1}^{K}\!\mathbb{E}\left[A_k(t)^2+\nu_k(t)^2|\boldsymbol{\Theta}(t)\right] \\\vspace{-0.05cm}
&-\eta\sum_{k=1}^{K}\mathbb{E}[Y_k(t)\big(A_k(t)-\nu_k(t))|\boldsymbol{\Theta}(t)].
\end{split} 
\end{equation}
From the system model and the constraints in \eqref{cons1} and \eqref{cons2}, we have $\mathbb{E}[A_k^2(t)]=r_k^2\leq \lambda_k^2$, $\mathbb{E}[\nu_k^2(t)]\leq r_k^2 $, and $\mathbb{E}[R_k(t)^2]\leq R_{k,\max}^2$, where $R_{k,\max}$ is the maximum achievable rate of user $k$ (the rate that user $k$ has when it transmits with full power and all other users use zero power). Then we have
\begin{equation}
\begin{split}
&\Delta\big(\boldsymbol{\Theta}(t)\big) \leq\frac{1}{2}\sum\limits_{k=1}^{K}R_{k,\max}^2+\frac{2\eta +1}{2}\sum\limits_{k=1}^{K}\lambda_k^2\\
&-\!\!\sum\limits_{k=1}^{K}\mathbb{E}[Q_k(t)R_k(t)|\boldsymbol{\Theta}(t)]\!-\!\!\sum\limits_{k=1}^{K}\mathbb{E}[A_k(t)\big(\eta Y_k(t)-Q_k(t))|\boldsymbol{\Theta}(t)]\\
&+\eta\sum\limits_{k=1}^{K}\mathbb{E}[Y_k(t)\nu_k(t)|\boldsymbol{\Theta}(t)].\nonumber
\end{split}
\end{equation}
Adding the penalty $-V \mathbb{E}[f(\boldsymbol{\nu})|\boldsymbol{\Theta}(t)]$ to the one-step drift, after re-arranging the terms, we obtain Lemma~\ref{lemma1}.


\bibliographystyle{IEEEtran}
\end{document}